\documentclass[%
 aip,
amsmath,amssymb,
reprint,
]{revtex4-1}

\usepackage{graphicx}
\usepackage{dcolumn}
\usepackage{bm}

\usepackage[utf8]{inputenc}
\usepackage[T1]{fontenc}
\usepackage{mathptmx}
\usepackage{etoolbox}

\usepackage[hidelinks]{hyperref} 
\hypersetup{
 setpagesize=false,
 bookmarksnumbered=true,%
 bookmarksopen=true,%
 colorlinks=true,%
 linkcolor=blue,
 citecolor=blue,
}

\usepackage{amsmath}
\usepackage{amssymb}
\usepackage{bm}
\usepackage{tabularx}
\usepackage{multirow}
\usepackage{booktabs} 

\usepackage{braket}
\usepackage{times,multirow,amsfonts,bm,xspace,pifont}
\usepackage{soul}
\usepackage{ulem}
\usepackage{mathrsfs}
\usepackage{array}

\makeatletter
\def\@email#1#2{%
 \endgroup
 \patchcmd{\titleblock@produce}
  {\frontmatter@RRAPformat}
  {\frontmatter@RRAPformat{\produce@RRAP{*#1\href{mailto:#2}{#2}}}\frontmatter@RRAPformat}
  {}{}
}%
\makeatother
\begin{document}

\preprint{AIP/123-QED}

\title
{Quantum geometry in correlated electron phases: from flat band to dispersive band}
\author{Taisei Kitamura}
 \email[]{taisei-kitamura@g.ecc.u-tokyo.ac.jp}
 \affiliation{Department of Physics, Graduate School of Science, University of Tokyo, Tokyo 113-0033, Japan}
 \affiliation{RIKEN Center for Emergent Matter Science (CEMS), Wako 351-0198, Japan}

\author{Akito Daido}%
\affiliation{Department of Physics, Graduate School of Science, Kyoto University, Kyoto 606-8502, Japan}%

\author{Youichi Yanase}
\affiliation{Department of Physics, Graduate School of Science, Kyoto University, Kyoto 606-8502, Japan}%

\date{\today}

\begin{abstract}
Quantum geometry, describing the geometric properties of the Bloch wave function in momentum space, has recently been recognized as a fundamental concept in condensed matter physics. The flat-band system offers the paradigmatic platform where quantum geometry plays the essential role in correlated electron phases.
However, systems that suffer from significant effects of quantum geometry are not limited to flat-band systems; dispersive-band systems also exhibit quantum condensed phases driven by quantum geometry. 
In this perspective, we provide a transparent account of quantum geometry and its role in correlated electron phases, throughout flat-band and dispersive-band systems.

\end{abstract}

\maketitle

\section{Introduction}
The Hubbard model, which describes Bloch electrons experiencing many-body electron-electron interactions, is a minimal framework for studying correlated electron phases in solids.
In its simplest form, the Bloch electrons are modeled by a single band, while the many-body interaction is represented by an on-site Coulomb repulsion. The single band Hubbard model has been suggested as a useful tool for describing various phenomena in condensed matter physics such as superconductivity and magnetism~\cite{Mahan2010,Altland2010}.
By contrast, correlated electron systems with nontrivial band structures have been intensively studied in modern condensed matter physics.
Once the assumption of a "single-band model" is removed, Bloch electrons acquire nontrivial properties originating from their wave functions, and the topology of quantum matter becomes manifest~\cite{Hasan2010,Qi2011}.
Stimulated by the discovery of topological quantum phases, the importance of the Bloch wave function among a variety of phenomena in solid state systems has been recognized. For example, in research of transport phenomena, the geometric structure of the Bloch wave function in momentum space including the Berry curvature and the shift vector has been investigated since the 1980's~\cite{Nagaosa2010,Sodemann2015,Nagaosa2017}. This may be regarded as an early stage research of what we refer to here as ``quantum geometry.''

Early studies of quantum geometry primarily focused on phenomena in which many-body interactions play a subdominant role, and a variety of transport responses triggered by quantum geometry has been identified~\cite{Liu2025,Gao2025}. However, the physics of correlated electron systems driven by many-body interactions can also be influenced by the nontrivial properties of the Bloch wave function. 
For instance, in iron-based superconductors, antiferromagnetic fluctuations arising from nesting between disconnected Fermi surfaces require strong hybridization among the $d$ electrons. This can be understood as a kind of quantum-geometric effects, since a nontrivial feature of the Bloch wave function is indispensable~\cite{Kuroki2008}. Moreover, in flat-band ferromagnetism, it is known that the overlap of Wannier functions is crucial for the uniqueness of the ground state~\cite{Tasaki2020}; since the Fourier transform of a Wannier function is the Bloch wave function, this provides another example in which the nontriviality of the Bloch wave function plays an essential role.

These specific examples illustrate the necessity of a theoretical framework that enables a unified understanding of correlated electron phases emerging from the nontrivial geometric properties of the Bloch wave function in momentum space. This issue is solved by \textit{quantum geometry}~\cite{Resta2011,Torma2023}. In fact, the two examples above can be understood within this framework~\cite{Kitamura2025,Oh2025}.
Such a unified description utilizing the concept of quantum geometry not only allows for a deeper understanding of the established phenomena but also predicts the correlated electron phases driven by the nontrivial Bloch wave functions.
Indeed, on the one hand, it has recently been recognized that flat-band systems offer a paradigmatic platform for studies of quantum geometry in correlated electron phases~\cite{Rossi2021,Torma2022,Peotta2023,Yu2025,Liu2024}.
On the other hand, correlated electron phases in dispersive-band systems are also a growing research area of quantum geometry.

In this perspective, we propose a transparent picture of how Bloch electrons manifest quantum geometry and how quantum geometry underlies correlated electron phases, which encompasses systems ranging from flat bands to dispersive bands.
We begin by reviewing quantum geometry in the context of correlated electron phases, focusing on flat-band systems. 
We then demonstrate that correlated electron phases in flat-band systems have counterparts in dispersive-band systems with nontrivial quantum geometry.
In this way, we clarify the role of nontrivial quantum geometry in correlated phases of both flat-band and dispersive-band electron systems and elucidate the connection between these two regimes.

\section{Quantum geometry in correlated electron phases of flat-band systems~\label{sec:quantum_geometry}}

The properties of Bloch electrons are described by the following eigenvalue equation:
\begin{eqnarray}
    H(\bm{k}) \ket{u_n(\bm{k})} = \epsilon_n(\bm{k}) \ket{u_n(\bm{k})} \, .
\end{eqnarray}
Here, $H(\bm{k})$ denotes the single-particle Hamiltonian in matrix representation at momentum $\bm{k}$. The energy dispersion $\epsilon_n(\bm{k})$ together with the Bloch wave function%
\footnote{Strictly speaking, $\ket{u_n(\bm{k})}$ is the periodic part of the Bloch wave function $e^{i\bm{k}\cdot\bm{r}}\ket{u_n(\bm{k})}$. In what follows, however, we simply refer to $\ket{u_n(\bm{k})}$ as the Bloch wave function.}
$\ket{u_n(\bm{k})}$ characterize the properties of the Bloch electrons.

\begin{figure}[tbp]
  \includegraphics[width=1.0\linewidth]{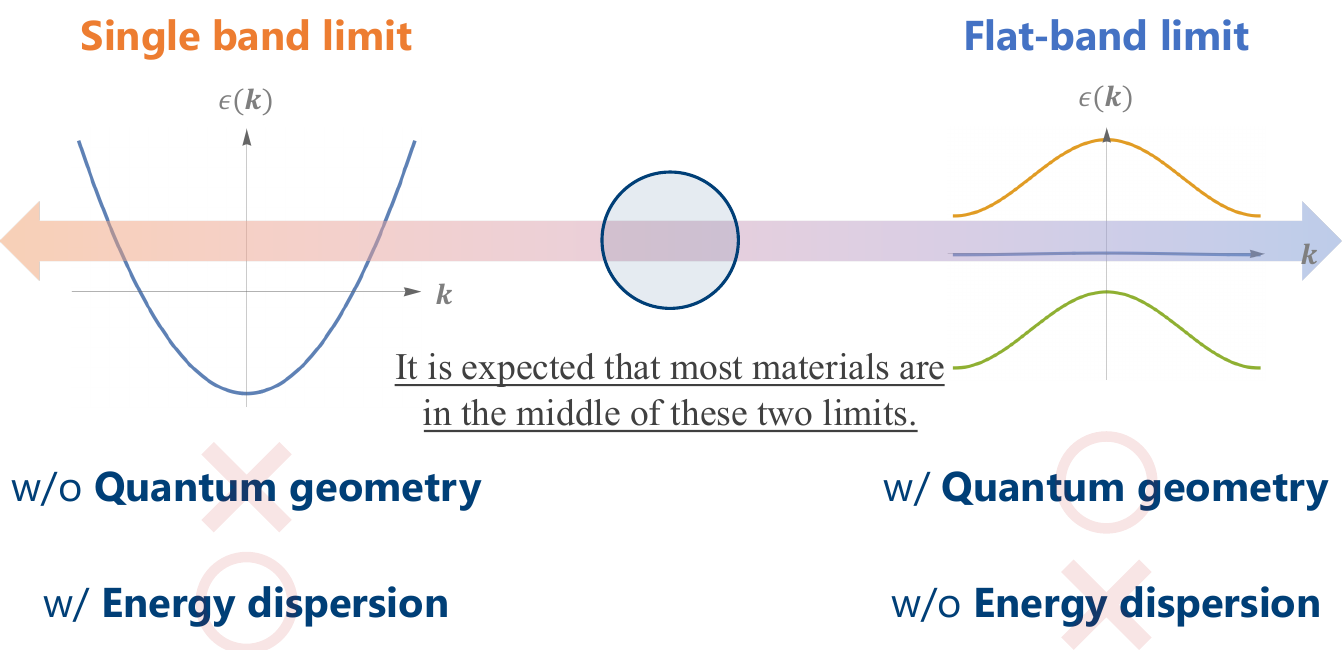}
  \centering
  \caption{Schematic illustration of the two limiting regimes of Bloch electrons.
 The left panel represents the single-band limit, and the right panel represents the flat-band limit.
 Most real materials are expected to lie between these two extremes.
}
  \label{fig:schematic}
\end{figure}

Focusing on these two quantities and assuming that only a single band lies near the Fermi energy, one can identify two limiting regimes for Bloch electrons (see Fig.~\ref{fig:schematic}).
One is the single-band limit, in which the band near the Fermi level is constructed by a single orbital that does not hybridize with other orbitals.
In this limit, only the energy dispersion depends on the momentum, while the Bloch wave function is independent of $\bm{k}$.
Consequently, physical phenomena are governed by the energy dispersion alone.
Indeed, in a standard solid-state textbook~\cite{Mahan2010}, a variety of topics in condensed matter physics are explained within a single-band picture by assuming, for example, a parabolic energy dispersion $\epsilon_n(\bm{k}) = k^2/2m$.

The other is the flat-band limit, where the one band in the vicinity of the Fermi energy becomes flat, i.e., $\bm{k}$-independent. Such a flat band originates from localized states generated by the destructive interference of Bloch wave functions due to orbital hybridization, and it appears in a variety of tight-binding models, including those on line graphs~\cite{Mielke1991,Tasaki2020}.
In sharp contrast to the single-band limit, only the Bloch wave function acquires a $\bm{k}$-dependence in this regime, and thus one expects physical phenomena to be characterized solely by the Bloch wave function. Although conventional wisdom has long considered many materials to effectively belong to the single-band limit, recent realizations of flat-band systems exemplified by twisted multilayer systems~\cite{Andrei2021,Nuckolls2024} have brought growing recognition of the importance of studying the flat-band limit.

Our interest here is to elucidate how the Bloch wave function characterizes physical phenomena in the flat-band limit.
For comparison, we first discuss how the energy dispersion characterizes phenomena in the conventional single-band limit, with focus on superconductivity and magnetism.

\subsection{Single-band limit and energy dispersion}
We consider the case in which the $n$th band constructing the Fermi surface belongs to the single-band limit.
In this situation, the Bloch wave function is identically unity, i.e., $\ket{u_n(\bm{k})}=1$.
First, we discuss superconductivity in this limit. Its hallmark phenomena are dissipationless current and the Meissner effect, which are described by the London equation,
\begin{eqnarray}
    j_\mu = D^{\rm s}_{\mu\nu}\, A_\nu \, ,
\end{eqnarray}
where $j_\mu$ and $A_\nu$ denote the electric current and the vector potential along $\mu,\nu=x,y,z$, respectively, and $D^{\rm s}_{\mu\nu}$ is the superfluid weight.
This equation indicates that superconducting responses emerge when the superfluid weight is finite.

Assuming fully gapped $s$-wave superconductivity for simplicity, the superfluid weight in the single-band limit is given by~\cite{Scalapino1992,Scalapino1993,Jujo2001},
\begin{eqnarray}
    D^{\rm s}_{\mu\nu} = \dfrac{2}{N}\sum_{\bm{k}} \frac{\nu^*_n(\bm{k})}{m^{\mu\nu}_n(\bm{k)}} \, ,
\end{eqnarray}
at sufficiently low temperatures within Fermi-liquid theory.
Here, $\nu^*_n(\bm{k}) =( 1 - \epsilon_n(\bm{k})/E_n(\bm{k}))/2$ is the zero-temperature expectation value of the electron number in the superconducting state at the momentum $\bm{k}$,
$E_n(\bm{k})=\sqrt{\epsilon_n^2(\bm{k})+\Delta^2}$ is the Bogoliubov quasiparticle spectrum, and $\Delta$ is the superconducting gap amplitude.
The inverse effective-mass tensor is determined solely by the energy dispersion as $[m^{\mu\nu}_n(\bm{k})]^{-1}=\partial_{k_\mu}\partial_{k_\nu}\epsilon_n(\bm{k})$.
This expression shows that in the single-band limit, the superfluid weight is determined by the effective mass of Bloch electrons and the momentum distribution function.

We next turn to itinerant magnetism.
Within the mean-field approximation, the magnetic order with wave vector $\bm{q}$ develops when the Stoner criterion~\cite{Hubbard1963}
\begin{eqnarray}
    U\,\chi_0(\bm{q}) \ge 1,
\end{eqnarray}
is satisfied with the on-site Coulomb interaction $U$.
In the single-band limit, the noninteracting magnetic susceptibility reads
\begin{eqnarray}
    \chi_0(\bm{q})
    = \frac{1}{N} \sum_{\bm{k}}
      \frac{f\!\bigl(\epsilon_n(\bm{k}+\bm{q})\bigr)-f\!\bigl(\epsilon_n(\bm{k})\bigr)}
           {\epsilon_n(\bm{k})-\epsilon_n(\bm{k}+\bm{q})} \, ,
\end{eqnarray}
whose right-hand side is the Lindhard function.
Here, $f(x)$ is the Fermi-Dirac distribution function and $N$ is the total number of unit cells.
In this formula, the dominant contribution arises from momenta near the Fermi surface satisfying $\epsilon_n(\bm{k})\!\sim\!-\epsilon_n(\bm{k}+\bm{q})$~\footnote{As in iron-based superconductors~\cite{Kuroki2008,Mazin2008}, nesting between electron and hole Fermi pockets can induce a $\log T$ divergence in susceptibility, and the corresponding nesting vectors connect momenta at which the velocities carry opposite signs.}.
For a "nested" structure of energy dispersion, such as those realized in cuprate superconductors~\cite{Yanase2003} 
,
this dominant contribution leads to divergence of susceptibility. In such cases, the nesting vector $\bm{q}$ that connects two momenta $\bm{k}$ and $\bm{k}+\bm{q}$ determines the stable magnetic structure.
For example, in cuprate high-$T_{\rm c}$ superconductors, nesting between segments of the Fermi surface with a large density of states divergently enhances magnetic susceptibility and leads to antiferromagnetic order with ${\bm q} \simeq (\pi,\pi)$~\cite{Yanase2003}.

Above examples illustrate that, in the single-band limit, the shape of the energy dispersion through, for example, the effective mass and nesting vector govern the physics of correlated electron phases.
This is the standard textbook knowledge of condensed matter physics~\cite{Mahan2010,Altland2010}.
By analogy, we can expect that in the other extreme regime, namely the flat-band limit, it is the shape of the Bloch wave function that characterizes physical phenomena.
We elaborate on this point in the next subsection.

\subsection{Flat-band limit and Bloch wave function~\label{sec:flat}}
Let us consider what is meant by the “shape of the Bloch wave function”.
Because the energy dispersion is a real scalar function, its profile can be visualized directly; thus, it is evident that its curvature (the effective mass) and the constant energy contour (the nesting of the Fermi surface) characterize the “shape of the energy dispersion”.
By contrast, the Bloch wave function is a complex vector, and its shape is not as easily visualized as energy dispersion.
Nevertheless, mathematics provides general frameworks to describe the “shape” of complex objects that are not real scalars; notable examples are topology and differential geometry.
We can apply these frameworks to Bloch wave functions and have obtained a theoretical tool to study condensed matter physics. This is the concept of \textit{quantum geometry}.
In this sense, quantum geometry includes topology.
The quantum geometry concerns the geometry of quantum mechanical wave functions in general. The parameters that define the geometry are not limited to the single-particle momentum, but in condensed matter physics the geometry of Bloch electrons defined in momentum space is considered in most cases.

While it has become widely appreciated that topology enables the understanding of various physical phenomena, it has recently become recognized that concepts of differential geometry applied to the Bloch wave function are useful for studying correlated electron phases~\cite{Rossi2021,Torma2022,Torma2023,Yu2025}.
A representative quantity is the Hilbert-Schmidt quantum distance:
\begin{eqnarray}
    D_{nn}^2(\bm{k},\bm{q})
    = 1 - \bigl|\braket{u_n(\bm{k}+\bm{q}) \vert u_n(\bm{k})}\bigr|^2 \, .
\end{eqnarray}
This quantity measures the distance between two Bloch states at $\bm{k}+\bm{q}$ and $\bm{k}$:
it vanishes when the two states are identical and reaches unity when they are completely different (orthogonal).
For small $\bm{q}$, it admits the expansion
\begin{eqnarray}
    D_{nn}^2(\bm{k},\bm{q}) \sim \sum_{\mu\nu} g_n^{\mu\nu}(\bm{k})\, q_\mu q_\nu \, ,
\end{eqnarray}
where
\begin{eqnarray}
    g_n^{\mu\nu}(\bm{k})
    = \sum_{m(\neq n)}
      \dfrac{
        \braket{\partial_{k_\mu} u_n(\bm{k}) \vert u_m(\bm{k})}
        \braket{u_m(\bm{k}) \vert \partial_{k_\nu} u_n(\bm{k})}
        + \mathrm{c.c.}
      }{2},
\end{eqnarray}
is the quantum metric~\cite{Provost1980,Resta2011}.
This metric encodes the infinitesimal distance between neighboring quantum states.

We should note that the Berry curvature, which characterizes the phase distance~\cite{Torma2023,Torma2022}, is considered to be an effective magnetic field in momentum space and is known to cause various topological phenomena~\cite{Nagaosa2010}.
In parallel, there is a clear physical interpretation of the quantum metric: it is the gauge-invariant part of the Wannier spread~\cite{Marzari1997,Vanderbilt2018}, a microscopic length scale of the electronic state. Therefore, the quantum metric also naturally plays essential roles in a variety of physical phenomena.
The Berry curvature and the quantum metric are related to each other through
the quantum geometric tensor 
\begin{align}
\mathscr{G}_{n}^{\mu\nu}(\bm k) = &
\braket{\partial_\mu u_n(\bm k)\vert\partial_\nu u_n(\bm k)} 
\notag \\ &
- \braket{\partial_\mu u_n(\bm k)\vert u_n(\bm k)}\braket{ u_n(\bm k)\vert\partial_\nu u_n(\bm k)},
\end{align}
whose real and imaginary parts are the quantum metric and the Berry curvature, respectively~\cite{Provost1980,Resta2011}. Because the quantum geometric tensor describes the geometric properties of Bloch electrons in momentum space, it is an essential quantity in the study of quantum geometry.

Let us first turn to the superfluid weight. Assuming fully gapped $s$-wave superconductivity again, the superfluid weight in the flat-band limit is given by~\cite{Peotta2015,Liang2017},
\begin{eqnarray}
    D_{\mu\nu}^{\rm s}
    = \frac{8}{N}\, |\Delta| \sqrt{\nu_n^*(1-\nu_n^*)}
      \sum_{\bm{k}} g_n^{\mu\nu}(\bm{k})
      \, , \label{eq:Ds_flat}
\end{eqnarray}
where we have used the fact that $\nu_n^*$ is $\bm{k}$-independent in a flat-band system.
Thus, in the flat-band limit, a finite quantum metric yields superconductivity.
This formula indicates that in flat-band systems, where electrons in real space are localized, the supercurrent is carried via the overlap of Wannier functions, producing a finite superconducting response such as the zero resistivity and the Meissner effect.
In the flat-band superconductivity of twisted-multilayer graphene~\cite{Cao2018,Hao2021,Park2021}, the theories~\cite{Hu2019,Xie2020,Julku2020,Hirobe2025} and the experiments~\cite{Tian2023,Tanaka2025,Banerjee2025} indicate the quantum geometric origin of the superfluid weight.
In certain topologically nontrivial flat-band systems, Eq.~\eqref{eq:Ds_flat} is bounded from below by topological invariants including the Chern number; in such cases, superconductivity is ensured by topology~\cite{Peotta2015,Xie2020,Herzog2022prl,Yu2023,Prijon2025}.

Note that Eq.~\eqref{eq:Ds_flat} itself is, strictly speaking, an incomplete formula in systems with sublattice degrees of freedom~\cite{Huhtinen2022-mu}. This problem arises from the fact that the quantum metric depends on the positions of the sublattices while the superfluid weight should not in an explicit way. This issue can be resolved by taking the vertex correction into account or by replacing the quantum metric with the one minimized among all possible choices of the sublattice coordinates, namely the minimal quantum metric~\cite{Huhtinen2022-mu}. Practically, on the other hand, sublattices are often located on high-symmetry positions within a unit cell, and the quantum metric as naturally defined coincides with the minimal quantum metric. As a result, Eq.~\eqref{eq:Ds_flat} is valid in many cases, and its application is sometimes found in the literature.

Magnetism likewise reflects geometric structure in the flat-band limit,
where the noninteracting spin susceptibility reads,
\begin{eqnarray}
    \chi_0(\bm{q})
    = \frac{1}{4 T N} \sum_{\bm{k}} \bigl[ 1 - D_{nn}^2(\bm{k},\bm{q}) \bigr],
\label{eq:flat-band_chi}
\end{eqnarray}
with temperature $T$.
This formula indicates that magnetic order is determined not by Fermi-surface nesting but by the quantum distance.
As the quantum distance also stabilizes the flat-band Bose-Einstein condensation~\cite{Julku2021,Julku2021excitations}, its key role in stabilizing the ordered phase of flat-band systems is ubiquitous and not limited to magnetism.
In particular, near $\bm{q}=0$, we can expand~\cite{Kitamura2024},
\begin{eqnarray}
    \chi_0(\bm{q})
    \simeq \chi_0(0)
      - \frac{1}{4 T N}
        \sum_{\mu\nu} \sum_{\bm{k}}
        g_n^{\mu\nu}(\bm{k})\, q_\mu q_\nu
      + \cdots \, , \label{eq:chi0_flat}
\end{eqnarray}
where the quantum metric suppresses the susceptibility at finite $\bm{q}$.
Therefore, we find that the quantum metric induces ferromagnetic correlations in flat-band systems.
It was also pointed out that quantum geometry is related to the mechanism of other types of flat-band magnetism through the concept of quantum geometric nesting~\cite{Han2024,Zhang2025,Sun2025-li}.

The ferromagnetic instability from quantum metric is closely related to the so-called flat-band ferromagnetism~\cite{Mielke1991,Mielke1992,Tasaki1992,Mielke1993,Tasaki1995,Mielke1999,Katsura2010,Tasaki2020,Wu2020,Kang2024,Herzog2022,Pichler2025}, where rigorous results are available~\cite{Mielke1991,Mielke1992,Tasaki1992,Mielke1993,Tasaki1995,Mielke1999,Katsura2010,Tasaki2020}.
The uniqueness of the ground state is ensured by quantum geometry as follows~\cite{Kitamura2025}:
For the uniqueness of the ground state in flat-band ferromagnetism, an effective repulsive interaction between localized electrons in real space, which is mediated by the overlap of Wannier functions, is crucial~\cite{Tasaki2020}.
Recalling again that the gauge-invariant part of the Wannier spread is the quantum metric, the quantum metric plays a central role in flat-band ferromagnetism as well.
While this understanding is based on the analysis of instability from the paramagnetic phase, quantum geometry also plays a crucial role for the stability of flat-band ferromagnetism from the viewpoint of the spin stiffness~\cite{Wu2020,Kang2024,Herzog2022}.
In this way, the rigorous theory and the Stoner-type picture of ferromagnetism are connected through the quantum metric.

To summarize, in flat-band systems, the “shape” of the Bloch wave function, represented by the quantum metric, characterizes a wide range of physical phenomena, providing a unifying and transparent view.
Indeed, qualitative behaviors of flat-band systems are dominated by quantum geometry.
For example, the low-temperature scaling law of the superfluid weight arising from the quantum metric is markedly different from the conventional one anticipated from the Fermi-liquid theory~\cite{Hirobe2025,buthenhoff2025lowtemperaturescalinglawsunconventional,Penttila2025-fn}.
As the temperature dependence of the superfluid weight is a standard method for probing pairing symmetry~\cite{Hardy1993-sq}, the scaling law of the superfluid weight is expected to be useful in determining the symmetry of flat-band superconductivity.
Indeed, in twisted bilayer and trilayer graphene, which are representative flat-band systems, anomalous temperature dependences of the superfluid weight have been observed~\cite{Tanaka2025,Banerjee2025}, and it can be naturally explained in terms of the quantum-geometric effect~\cite{Hirobe2025}.

\section{From flat band to dispersive band~\label{sec:superconductivity}}
In the previous section, we discussed the two limiting cases, namely, the single-band limit and the flat-band limit. 
In the latter, the "shape" of the Bloch wave function dominates the physical phenomena, while the energy dispersion plays a central role in the former. The twisted bilayer graphene~\cite{Cao2018} and cuprate superconductors~\cite{Yanase2003} may be close to the two limiting cases.
However, it is expected that many materials lie in the intermediate regime between these limits.
Among them are numerous systems with nontrivial quantum geometry arising from band degeneracies such as topological materials~\cite{Vergniory2019,Zhang2019}, which may host unconventional phenomena unexpected from the analysis of band dispersion. 
In other words, dispersive-band systems can also possess nontrivial quantum geometry, driving various correlated phases.
Recently, this direction has seen significant developments: various superconducting~\cite{Ahn2021,Chen2021,Kitamura2022,Kitamura2023,Kitamura2024,Yu2025,Daido2024,Zhang2024,Simon2025} and magnetic~\cite{Kitamura2024,Heinsdorf2025,Kitamura2025,Kudo2025,Hu2025,Oh2025,Espinosachampo2025,Wang2025,Nogaki2025,shimizu2026magnetic} phenomena in dispersive-band systems have been identified as being driven by quantum geometry.
In particular, the quantum-geometry-driven superconducting and magnetic phenomena in flat-band systems discussed in Sec.~\ref{sec:flat} carry over to electrons in certain dispersive band systems.

First, we consider the superfluid weight of multi-band systems.
Assuming isotropic $s$-wave pairing, the superfluid weight is given by~\cite{Liang2017,Kitamura2022superconductivity},
\begin{align}
  D_{\mu\nu}^{\rm s} =& D_{\mu\nu}^{\rm conv} + D_{\mu\nu}^{\rm geom} \, , \\
  D_{\mu\nu}^{\rm conv}
  =& \dfrac{2}{N}\sum_{n,\bm{k}}\!\left[
      \frac{\nu_n^*(\bm{k})}{m_{n}^{\mu\nu}(\bm{k})}\tanh\!\bigl(E_n(\bm{k})/2T\bigr)
      \right.
      \notag\\
      &+\left.
       \,\nu_n^*(\bm{k})\, v_n^\mu(\bm{k})\, v_n^\nu(\bm{k})\,
        f'\!\bigl(E_n(\bm{k})\bigr)
    \right] , \label{eq:Ds_conv}\\
  D_{\mu\nu}^{\rm geom}
  =& 8\dfrac{|\Delta|}{N}\sum_{n\neq m,\bm{k}}
      \sqrt{\nu_{n}^*(\bm k)(1-\nu_{n}^*(\bm k))}\,
      \frac{\epsilon_{m}(\bm{k})-\epsilon_{n}(\bm{k})}{\epsilon_{n}(\bm{k})+\epsilon_{m}(\bm{k})}\,\notag\\
      &\times
      g_{nm}^{\mu\nu}(\bm{k})\,
      \tanh\!\bigl(E_{n}(\bm{k})/2T\bigr) ,
\end{align}
where $g_{nm}^{\mu\nu}(\bm{k})
= \braket{\partial_{k_\mu}u_n(\bm{k})\vert u_m(\bm{k})}
  \braket{u_m(\bm{k})\vert\partial_{k_\nu}u_n(\bm{k})}/2 + \mathrm{c.c.}$
is the band-resolved quantum metric, and
$v_n^\mu(\bm{k}) = \partial_{k_\mu}\epsilon_n(\bm{k})$ denotes the group velocity.
In the single-band limit only the Fermi-liquid contribution $D_{\mu\nu}^{\rm conv}$ is finite,
whereas in the flat-band limit, only the quantum-geometric contribution
$D_{\mu\nu}^{\rm geom}$ remains finite.
Note that, similarly  to the case of Eq.~\eqref{eq:Ds_flat}, these formulas 
are also affected by the internal position of sublattices. This problem is resolved in a similar manner to the case of Eq.~\eqref{eq:Ds_flat}~\cite{Huhtinen2022-mu}.

The ratio $D_{\mu\nu}^{\rm geom}/D_{\mu\nu}^{\rm conv}$ is typically of the order $|\Delta|/E_{\rm F}$ with the Fermi energy $E_{\rm F}$; hence, in ordinary multiband superconductors, the quantum-geometric contribution is much smaller than the Fermi-liquid contribution and has often been neglected.
However, in the Bardeen-Cooper-Schrieffer (BCS)-Bose-Einstein condensation (BEC) crossover regime~\cite{Nozieres1985}, $|\Delta|/E_{\rm F}$ is in the order of unity, which implies that the quantum geometric contribution is comparable to the conventional Fermi-liquid contribution.
Indeed, monolayer FeSe does not have a flat band, but is near the
BCS-BEC crossover regime~\cite{Kasahara2014,Kasahara2016,Hanaguri2019,Kasahara2020} and hosts nontrivial quantum geometry due to the band degeneracies near the Fermi level~\cite{Kitamura2021}. The quantum geometric contribution to the superfluid weight has been revealed to be comparable to the conventional term~\cite{Kitamura2022superconductivity}.
Moreover, it has also been shown that, in certain parameter sets, quantum geometry enhances the Berezinskii-Kosterlitz-Thouless transition temperature~\cite{Kosterlitz1973, Berezinskii1971}, which corresponds to the superconducting transition temperature of two-dimensional (2D) systems, by nearly 10K via the superfluid weight.

Next, we focus on the itinerant magnetism in multi-band systems.
The noninteracting spin susceptibility of multiband systems~\footnote{Here we assume the absence of the spin-orbit coupling for simplicity.} can be written as
\begin{align}
    \chi_0(\bm{q})
    =& \frac{1}{N}\sum_{\bm{k}}\sum_{n m}
     \frac{f\!\bigl(\epsilon_n(\bm{k}+\bm{q})\bigr)-f\!\bigl(\epsilon_m(\bm{k})\bigr)}
           {\epsilon_m(\bm{k})-\epsilon_n(\bm{k}+\bm{q})}\, \notag\\
    &\times
      \Bigl[1 - D_{nm}^2(\bm{k},\bm{q})\Bigr] \, .
\end{align}
Here, $D_{nm}^2(\bm{k},\bm{q}) = 1 - \bigl|\braket{u_n(\bm{k}+\bm{q}) \vert u_m(\bm{k})}\bigr|^2$
is the quantum distance generalized to include off-diagonal interband components.
We see that the quantum distance appears in a similar manner to Eq.~\eqref{eq:flat-band_chi} for flat-band systems.
However, in contrast to the flat-band limit, the effect of the dispersive energy band appears as in the Lindhard function, which can reflect Fermi-surface nesting. Therefore, the contributions of the energy dispersion and quantum geometry are mixed in a complicated manner. 

Nevertheless, around $\bm{q}=0$ we obtain a transparent formula~\cite{Kitamura2024} as,
\begin{eqnarray}
    \chi_0(\bm{q})
    &=& \chi_0(0)
      + \sum_{\mu\nu}\!\left(\chi_{\rm geom}^{\mu\nu} + \chi_{\rm mass}^{\mu\nu}\right)
        q_\mu q_\nu
      + \cdots ,\\
    \chi_{\rm geom}^{\mu\nu}
    &=& \sum_{n}\int \frac{d\bm{k}}{(2\pi)^2}
        \big[
          2 f'\!\bigl(\epsilon_n(\bm{k})\bigr)\, g_n^{\mu\nu}(\bm{k})
          \notag\\
          &&+ 4 f\!\bigl(\epsilon_n(\bm{k})\bigr)\, X_n^{\mu\nu}(\bm{k})
        \big] ,\\
    \chi_{\rm mass}^{\mu\nu}
    &=& - \sum_{n}\int \frac{d\bm{k}}{(2\pi)^2}
        \frac{f^{(2)}\!\bigl(\epsilon_n(\bm{k})\bigr)}{6}\,
        \bigl[m_n^{\mu\nu}(\bm{k})\bigr]^{-1} ,
\end{eqnarray}
where the quadratic coefficient in $\bm{q}$ is decomposed as $\chi_{\rm c}^{\mu\nu} = \chi_{\rm geom}^{\mu\nu} + \chi_{\rm mass}^{\mu\nu}$.
Here, $\chi_{\rm geom}^{\mu\nu}$ is the quantum geometric term originating from the quantum
distance, which contains the quantum metric $g_n^{\mu\nu}(\bm{k})$ in addition to another geometric quantity
called positional shift (or Berry connection polarizability),
$X_n^{\mu\nu}(\bm{k}) = \sum_{m(\neq n)} g_{nm}^{\mu\nu}(\bm{k})/
\bigl(\epsilon_m(\bm{k})-\epsilon_n(\bm{k})\bigr)$~\cite{Gao2014}.
The second term $\chi_{\rm mass}^{\mu\nu}$ is the effective-mass contribution originating from the Lindhard function and is determined by the energy dispersion.

The quadratic coefficient $\chi_{\rm c}^{\mu\nu} = \chi_{\rm geom}^{\mu\nu} + \chi_{\rm mass}^{\mu\nu}$ represents the curvature of susceptibility and plays
a key role for examining ferromagnetism:
if $\chi_{\rm  c}^{\mu\nu}$ is positive semidefinite, $\chi_0(\bm{q})$ does not peak at $\bm{q}=0$ and ferromagnetism is prohibited. In contrast, a negative definite $\chi_{\rm c}^{\mu\nu}$ implies that $\chi_0(0)$
is a local maximum, indicating the presence of ferromagnetic correlations.
Thus, $\chi_{\rm c}$ serves as an indicator of ferromagnetism.
Analysis of this quantity provides a transparent view of the role of quantum geometry. 
In the quantum geometric term $\chi_{\rm geom}^{\mu\nu}$, the contribution
involving the positional shift is positive, whereas the contribution involving the quantum metric
is negative and thus promotes ferromagnetic correlations.
This is consistent with the discussion in Sec.~\ref{sec:flat}, where a close relation between the quantum metric and flat-band ferromagnetism is discussed. In other words, the concept of flat-band ferromagnetism can be generalized to dispersive band systems by utilizing the quantum metric.
Even when the effective-mass term disfavors ferromagnetism, the quantum geometric term can overcome it, resulting in \textit{quantum geometric ferromagnetism}.
Stimulated by this discovery, the competition between quantum geometric and energy-dispersion effects on the magnetism has attracted much attention~\cite{Heinsdorf2025,Kitamura2025,Kudo2025,Oh2025,Nogaki2025,shimizu2026magnetic}, and an approach based on the effective spin model has also been proposed recently~\cite{Hu2025}.

A typical platform for quantum geometric ferromagnetism is a 2D $t_{2g}$-orbital model with symmetry-protected band
degeneracies~\cite{Kitamura2025}.
It is shown that~\cite{Kitamura2025}, in most parameters where band touching lies on Fermi surfaces, the 2D $t_{2g}$-orbital model exhibits quantum geometric ferromagnetism.
The band-touching point in this model is accompanied by a saddle-point structure with a divergent density of states (DOS), which is the so-called singular saddle point. The divergent quantum metric induces the ferromagnetic correlation despite the effective mass being disadvantageous for it, and the divergent DOS ensures the Stoner criteria. Therefore, it is expected that the singular saddle point is a promising platform for quantum geometric ferromagnetism, which is not restricted to a specific model.

Moreover, in the same 2D $t_{2g}$-orbital model, a flat band emerges at a certain parameter set.
In this case, we can rigorously prove the ferromagnetic ground state by using the theory of flat-band ferromagnetism~\cite{Kitamura2025}.
Hence, quantum geometric ferromagnetism is continuously connected to flat-band ferromagnetism.
It follows that the essence of flat-band ferromagnetism is taken over to dispersive band systems through the
divergent quantum metric due to the band touching.

\section{Summary and outlook~\label{sec:summary}}
In summary, the properties of Bloch electrons are described by the energy spectrum and the Bloch wave function, whose shape in momentum space, that is, energy dispersion and quantum geometry, respectively, characterize various phenomena in condensed matter physics. In particular, Bloch electrons possess two characteristic quantum geometric structures: one is the phase distance described by the Berry curvature, and the other is the amplitude distance related to the quantum metric.
While the Berry curvature plays essential roles in various topological phenomena as an effective magnetic field, the quantum metric serves as the spread of the Wannier function, enriching a variety of correlated electron phases.

Flat-band systems without energy dispersion offer a paradigmatic platform for correlated electron phases dominated by quantum geometry and central topics for studying quantum geometry.
However, we have shown that the quantum geometric effect can be taken over to the dispersive-band systems without a flat band; various types of superconductivity and magnetism can be driven by quantum geometry.
While we focus on the superfluid weight and ferromagnetism, quantum geometry also plays a significant role in other superconducting and magnetic phenomena. For example, quantum geometry induces finite momentum superconductivity, including Fulde–Ferrell–Larkin–Ovchinnikov superconductivity~\cite{Kitamura2022} and anapole superconductivity~\cite{Kitamura2023}, which have flat-band counterparts~\cite{Jiang2023,Chen2023,Wang2025,Aaron2025}.
As for magnetism, recent progress has shown that the quantum-geometry-driven magnetism is not limited to ferromagnetsim: antiferromagnetism~\cite{Oh2025}, altermagnetism~\cite{Heinsdorf2025}, and odd-parity magnetism~\cite{Kudo2025} can be driven by quantum geometry.
Moreover, the quantum geometric origins of magnetism in the real materials, LaFeAsO and Pb$_9$Cu(PO$_4$)$_6$O, are revealed by the first-principles calculations~\cite{shimizu2026magnetic}.

Beyond the phenomena covered in this perspective, quantum geometry manifests itself in many other contexts.
In addition to well-known transport responses~\cite{Liu2025,Gao2025}, it plays roles in correlated electron physics, such as the electron-phonon coupling phenomena~\cite{Hu2021,Yu2024non,Hu2025ph1,Hu2025ph2,Pellitteri2025,Li2025}, fractional Chern insulators~\cite{Liu2024}, Bose-Einstein condensation~\cite{Julku2021,Julku2021excitations}, and screened Coulomb interaction~\cite{Shavit2025,Shavit2025nematic}.
As we have seen, even in problems traditionally thought to be governed primarily by the energy
dispersion, consideration of flat-band systems shows that the phenomena are necessarily controlled
by quantum geometry, and these effects carry over to dispersive-band systems as well.
We hope that this article will help catalyze further research on quantum geometry, leading to the
recognition of its importance across diverse phenomena and to the discovery of yet unrecognized physics.

\begin{acknowledgments}
We are grateful to T. Yamashita, J. Ishizuka, S. Kanasugi, M. Chazono, H. Nakai, C.G. Oh, J.W. Rhim, M. Ichikawa, M. Fujimoto, H. Hirobe, H. Katsura, R. Arita, and H. Watanabe for fruitful comments and discussions.
This work was supported by JSPS KAKENHI (Grant No. JP25K23367, No. JP18H01178, No. JP18H05227, No. JP20H05159, No. JP21K18145, No. JP22H01181, No. JP22H04933,  No. JP21K13880, No. JP23K17353, No. JP24H01662, No. JP23KK0248, No. JP24H00007, No. JP24K21530, No. JP25H01249).
\end{acknowledgments}

\section*{Data Availability Statement}
The data that support the findings of this study are available from the corresponding author upon reasonable request.

\appendix

\bibliography{aipsamp}

@misc{shimizu2026magnetic,
      title={Magnetic fluctuations driven by quantum geometry}, 
      author={Makoto Shimizu and Chang-guen Oh and Youichi Yanase},
      year={2026},
      eprint={2602.14511},
      archivePrefix={arXiv},
      primaryClass={cond-mat.str-el},
      url={https://arxiv.org/abs/2602.14511},
}

@ARTICLE{Penttila2025-fn,
  title     = "{Flat-band ratio and quantum metric in the superconductivity of
               modified Lieb lattices}",
  author    = "Penttilä, Reko P S and Huhtinen, Kukka-Emilia and Törmä, Päivi",
  journal   = "Commun. Phys.",
  publisher = "Springer Science and Business Media LLC",
  volume    =  8,
  number    =  1,
  pages     =  50,
  month     =  jan,
  year      =  2025,
  URL = "https://www.nature.com/articles/s42005-025-01964-y"
}

@misc{buthenhoff2025lowtemperaturescalinglawsunconventional,
      title={Low-temperature scaling laws in unconventional flat-band superconductors}, 
      author={Maximilian Buthenhoff and Yusuke Nishida},
      year={2025},
      eprint={2510.23159},
      archivePrefix={arXiv},
      primaryClass={cond-mat.supr-con},
      url={https://arxiv.org/abs/2510.23159}
}

@ARTICLE{Sun2025-li,
  title     = "{Flat-band Fulde-Ferrell-Larkin-Ovchinnikov state from quantum
               geometric discrepancy}",
  author    = "Sun, Zi-Ting and Yu, Ruo-Peng and Chen, Shuai A and Hu, Jin-Xin
               and Law, K T",
  journal   = "Quantum Front.",
  publisher = "Springer Science and Business Media LLC",
  volume    =  4,
  number    =  1,
  pages     =  20,
  month     =  dec,
  year      =  2025,
  URL = "https://link.springer.com/article/10.1007/s44214-025-00093-5"
}

@ARTICLE{Hu2025ph1,
  title     = "{Quantum Geometry in Phonon-Mediated Optical Responses}",
  author    = "Hu, Jiaming and Li, Wenbin and Guo, Zhichao and Wang, Hua and
               Chang, Kai",
  journal   = "Phys. Rev. Lett.",
  publisher = "American Physical Society (APS)",
  volume    =  135,
  number    =  25,
  pages     =  256404,
  month     =  dec,
  year      =  2025,
  URL = "https://journals.aps.org/prl/abstract/10.1103/y66p-kjj7"
}

@ARTICLE{Hu2025ph2,
  title     = "{Quantum Geometric Origin of Strain-Induced Ferroelectric Phase
               Transitions}",
  author    = "Hu, Jiaming and Zhu, Ziye and Yuan, Yubo and Wang, Hua and Chang,
               Kai",
  journal   = "Phys. Rev. Lett.",
  publisher = "American Physical Society (APS)",
  volume    =  135,
  number    =  25,
  pages     =  256405,
  month     =  dec,
  year      =  2025,
  URL = "https://journals.aps.org/prl/abstract/10.1103/dbxd-jk76"
}

@misc{Li2025,
      title={Phonon Dichroisms Revealing Unusual Electronic Quantum Geometry}, 
      author={Ding Li and Guoao Yang and Tao Qin and Jianhui Zhou and Yugui Yao},
      year={2025},
      eprint={2511.16141},
      archivePrefix={arXiv},
      primaryClass={cond-mat.mes-hall},
      url={https://arxiv.org/abs/2511.16141}, 
}

@misc{Pichler2025,
      title={False Vacuum Decay in Flat-Band Ferromagnets: Role of Quantum Geometry and Chiral Edge States}, 
      author={Fabian Pichler and Clemens Kuhlenkamp and Michael Knap},
      year={2025},
      eprint={2512.13786},
      archivePrefix={arXiv},
      primaryClass={cond-mat.str-el},
      url={https://arxiv.org/abs/2512.13786}
}

@ARTICLE{Simon2025,
  title     = "Normal-state quantum geometry, nonlocality, and superconductivity",
  author    = "Simon, Florian",
  journal   = "Phys. Rev. B.",
  publisher = "American Physical Society (APS)",
  volume    =  112,
  number    =  10,
  pages     =  104504,
  month     =  sep,
  year      =  2025,
  URL = "https://journals.aps.org/prb/abstract/10.1103/kzj9-99j1"
}

@ARTICLE{Zhang2024,
  title     = "{Quantum-Geometry-Induced Anomalous Hall Effect in Nonunitary Superconductors and Application to} {Sr}$_2${RuO}$_{4}$",
  author    = "Zhang, Jia-Long and Chen, Weipeng and Liu, Hao-Tian and Li, Yu
               and Wang, Zhiqiang and Huang, Wen",
  journal   = "Phys. Rev. Lett.",
  publisher = "American Physical Society (APS)",
  volume    =  132,
  number    =  13,
  pages     =  136001,
  month     =  mar,
  year      =  2024,
  URL = "https://journals.aps.org/prl/abstract/10.1103/PhysRevLett.132.136001"
}

@misc{Espinosachampo2025,
      title={{Magnetic phase transitions protected by topological quantum geometry transitions: effects of electron-electron interactions in the Creutz ladder system}}, 
      author={Abdiel de Jesús Espinosa-Champo and Gerardo G. Naumis},
      year={2025},
      eprint={2509.26320},
      archivePrefix={arXiv},
      primaryClass={cond-mat.str-el},
      url={https://arxiv.org/abs/2509.26320}, 
}

@misc{Pellitteri2025,
      title={{Phonon spectra, quantum geometry, and the Goldstone theorem}}, 
      author={Guglielmo Pellitteri and Zenan Dai and Haoyu Hu and Yi Jiang and Guido Menichetti and Andrea Tomadin and B. Andrei Bernevig and Marco Polini},
      year={2025},
      eprint={2502.04221},
      archivePrefix={arXiv},
      primaryClass={cond-mat.mes-hall},
      url={https://arxiv.org/abs/2502.04221}, 
}

@ARTICLE{Chen2021,
  title     = "Quantum-geometry-induced intrinsic optical anomaly in
               multiorbital superconductors",
  author    = "Chen, Weipeng and Huang, Wen",
  journal   = "Phys. Rev. Research",
  publisher = "American Physical Society",
  volume    =  3,
  number    =  4,
  pages     = "L042018",
  month     =  nov,
  year      =  2021,
  URL = "https://journals.aps.org/prresearch/abstract/10.1103/PhysRevResearch.3.L042018"
}

@ARTICLE{Ahn2021,
  title     = "Superconductivity-induced spectral weight transfer due to quantum
               geometry",
  author    = "Ahn, Junyeong and Nagaosa, Naoto",
  journal   = "Phys. Rev. B Condens. Matter",
  publisher = "American Physical Society",
  volume    =  104,
  number    =  10,
  pages     = "L100501",
  month     =  sep,
  year      =  2021,
  URL = "https://journals.aps.org/prb/abstract/10.1103/PhysRevB.104.L100501"
}

@article{Kasahara2020,
  title = {{Evidence for an Fulde-Ferrell-Larkin-Ovchinnikov State with Segmented Vortices in the BCS-BEC-Crossover Superconductor FeSe}},
  author = {Kasahara, S. and Sato, Y. and Licciardello, S. and \ifmmode \check{C}\else \v{C}\fi{}ulo, M. and Arsenijevi\ifmmode \acute{c}\else \'{c}\fi{}, S. and Ottenbros, T. and Tominaga, T. and B\"oker, J. and Eremin, I. and Shibauchi, T. and Wosnitza, J. and Hussey, N. E. and Matsuda, Y.},
  journal = {Phys. Rev. Lett.},
  volume = {124},
  issue = {10},
  pages = {107001},
  numpages = {6},
  year = {2020},
  month = {Mar},
  publisher = {American Physical Society},
  doi = {10.1103/PhysRevLett.124.107001},
  url = {https://link.aps.org/doi/10.1103/PhysRevLett.124.107001}
}

@article {Kasahara2014,
	author = {Kasahara, Shigeru and Watashige, Tatsuya and Hanaguri, Tetsuo and Kohsaka, Yuhki and Yamashita, Takuya and Shimoyama, Yusuke and Mizukami, Yuta and Endo, Ryota and Ikeda, Hiroaki and Aoyama, Kazushi and Terashima, Taichi and Uji, Shinya and Wolf, Thomas and von L{\"o}hneysen, Hilbert and Shibauchi, Takasada and Matsuda, Yuji},
	title = {{Field-induced superconducting phase of FeSe in the BCS-BEC cross-over}},
	volume = {111},
	number = {46},
	pages = {16309--16313},
	year = {2014},
	doi = {10.1073/pnas.1413477111},
	publisher = {National Academy of Sciences},
	issn = {0027-8424},
	URL = {https://www.pnas.org/content/111/46/16309},
	eprint = {https://www.pnas.org/content/111/46/16309.full.pdf},
	journal = {Proceedings of the National Academy of Sciences}
}

@article{Kasahara2016,
author={Kasahara, S. and Yamashita, T. and Shi, A. and Kobayashi, R. and Shimoyama, Y. and Watashige, T. and Ishida, K. and Terashima, T. and Wolf, T. and Hardy, F.
and Meingast, C. and L{\"o}hneysen, H. v. and Levchenko, A. and Shibauchi, T.and Matsuda, Y.},
title={{Giant superconducting fluctuations in the compensated semimetal FeSe at the BCS--BEC crossover}},
journal={Nature Communications},
year={2016},
month={Sep},
day={30},
volume={7},
number={1},
pages={12843},
issn={2041-1723},
doi={10.1038/ncomms12843},
url={https://doi.org/10.1038/ncomms12843}
}

@article{Hanaguri2019,
  title = {{Quantum Vortex Core and Missing Pseudogap in the Multiband BCS-BEC Crossover Superconductor FeSe}},
  author = {Hanaguri, T. and Kasahara, S. and B\"oker, J. and Eremin, I. and Shibauchi, T. and Matsuda, Y.},
  journal = {Phys. Rev. Lett.},
  volume = {122},
  issue = {7},
  pages = {077001},
  numpages = {5},
  year = {2019},
  month = {Feb},
  publisher = {American Physical Society},
  doi = {10.1103/PhysRevLett.122.077001},
  url = {https://link.aps.org/doi/10.1103/PhysRevLett.122.077001}
}

@Article{Nozieres1985,
author={Nozi{\`e}res, P.
and Schmitt-Rink, S.},
title={Bose condensation in an attractive fermion gas: From weak to strong coupling superconductivity},
journal={Journal of Low Temperature Physics},
year={1985},
month={May},
day={01},
volume={59},
number={3},
pages={195-211},
issn={1573-7357},
doi={10.1007/BF00683774},
url={https://doi.org/10.1007/BF00683774}
}

@ARTICLE{Hasan2010,
  title     = "Colloquium: Topological insulators",
  author    = "Hasan, M Z and Kane, C L",
  journal   = "Rev. Mod. Phys.",
  publisher = "American Physical Society",
  volume    =  82,
  number    =  4,
  pages     = "3045--3067",
  month     =  nov,
  year      =  2010,
  URL = "https://journals.aps.org/rmp/abstract/10.1103/RevModPhys.82.3045"
}

@ARTICLE{Qi2011,
  title     = "Topological insulators and superconductors",
  author    = "Qi, Xiao-Liang and Zhang, Shou-Cheng",
  journal   = "Rev. Mod. Phys.",
  publisher = "American Physical Society",
  volume    =  83,
  number    =  4,
  pages     = "1057--1110",
  month     =  oct,
  year      =  2011,
  URL = "https://journals.aps.org/rmp/abstract/10.1103/RevModPhys.83.1057"
}

@article{Julku2021,
  title = {{Q}uantum {G}eometry and {F}lat {B}and {B}ose-{E}instein {C}ondensation},
  author = {Julku, Aleksi and Bruun, Georg M. and T\"orm\"a, P\"aivi},
  journal = {Phys. Rev. Lett.},
  volume = {127},
  issue = {17},
  pages = {170404},
  numpages = {6},
  year = {2021},
  month = {Oct},
  publisher = {American Physical Society},
  doi = {10.1103/PhysRevLett.127.170404},
  url = {https://link.aps.org/doi/10.1103/PhysRevLett.127.170404}
}

@article{Julku2021excitations,
  title = {{E}xcitations of a {B}ose-{E}instein condensate and the quantum geometry of a flat band},
  author = {Julku, Aleksi and Bruun, Georg M. and T\"orm\"a, P\"aivi},
  journal = {Phys. Rev. B},
  volume = {104},
  issue = {14},
  pages = {144507},
  numpages = {15},
  year = {2021},
  month = {Oct},
  publisher = {American Physical Society},
  doi = {10.1103/PhysRevB.104.144507},
  url = {https://link.aps.org/doi/10.1103/PhysRevB.104.144507}
}

@Article{Yu2024non,
author={Yu, Jiabin
and Ciccarino, Christopher J.
and Bianco, Raffaello
and Errea, Ion
and Narang, Prineha
and Bernevig, B. Andrei},
title={Non-trivial quantum geometry and the strength of electron--phonon coupling},
journal={Nature Physics},
year={2024},
month={Aug},
day={01},
volume={20},
number={8},
pages={1262-1268},
issn={1745-2481},
doi={10.1038/s41567-024-02486-0},
url={https://doi.org/10.1038/s41567-024-02486-0}
}

@ARTICLE{Hu2021,
  title     = "{Phonon Helicity Induced by Electronic Berry Curvature in Dirac
               Materials}",
  author    = "Hu, Lun-Hui and Yu, Jiabin and Garate, Ion and Liu, Chao-Xing",
  journal   = "Phys. Rev. Lett.",
  publisher = "American Physical Society",
  volume    =  127,
  number    =  12,
  pages     =  125901,
  month     =  sep,
  year      =  2021,
  URL = "https://journals.aps.org/prl/abstract/10.1103/PhysRevLett.127.125901"
}

@ARTICLE{Liu2025,
  title     = "Quantum geometry in condensed matter",
  author    = "Liu, Tianyu and Qiang, Xiao-Bin and Lu, Hai-Zhou and Xie, X C",
  journal   = "Natl. Sci. Rev.",
  publisher = "Oxford University Press (OUP)",
  volume    =  12,
  number    =  3,
  pages     = "nwae334",
  month     =  mar,
  year      =  2025,
  URL = "https://academic.oup.com/nsr/article/12/3/nwae334/7762198?login=true"
}

@misc{Gao2025,
      title={Quantum Geometry Phenomena in Condensed Matter Systems}, 
      author={Anyuan Gao and Naoto Nagaosa and Ni Ni and Su-Yang Xu},
      year={2025},
      eprint={2508.00469},
      archivePrefix={arXiv},
      primaryClass={cond-mat.str-el},
      url={https://arxiv.org/abs/2508.00469}, 
}

@misc{Nogaki2025,
      title={Quantum geometric magnetic monopole and two-phase superconductivity in {CeRh}$_{2}${As}$_{2}$}, 
      author={Kosuke Nogaki and Youichi Yanase},
      year={2025},
      eprint={2510.24289},
      archivePrefix={arXiv},
      primaryClass={cond-mat.supr-con},
      url={https://arxiv.org/abs/2510.24289}, 
}

@misc{Hu2025,
      title={{Ferromagnetism vs. Antiferromagnetism in Narrow-Band Systems: Competition Between Quantum Geometry and Band Dispersion}}, 
      author={Haoyu Hu and Oskar Vafek and Kristjan Haule and B. Andrei Bernevig},
      year={2025},
      eprint={2509.03575},
      archivePrefix={arXiv},
      primaryClass={cond-mat.str-el},
      url={https://arxiv.org/abs/2509.03575}, 
}

@misc{Shavit2025nematic,
      title={{Nematic Enhancement of Superconductivity in Multilayer Graphene via Quantum Geometry}}, 
      author={Gal Shavit},
      year={2025},
      eprint={2509.13407},
      archivePrefix={arXiv},
      primaryClass={cond-mat.supr-con},
      url={https://arxiv.org/abs/2509.13407}, 
}

@ARTICLE{Shavit2025,
  title     = "{Quantum Geometric Kohn-Luttinger Superconductivity
}",
  author    = "Shavit, Gal and Alicea, Jason",
  journal   = "Phys. Rev. Lett.",
  publisher = "American Physical Society (APS)",
  volume    =  134,
  number    =  17,
  pages     =  176001,
  month     =  apr,
  year      =  2025,
  URL = "https://journals.aps.org/prl/abstract/10.1103/PhysRevLett.134.176001"
}

@misc{Aaron2025,
      title={Quantum Geometry of Time-Reversal Symmetry Breaking in Flat-Band Superconductivity}, 
      author={Aaron Dunbrack and Pauli Virtanen and Tero T. Heikkilä},
      year={2025},
      eprint={2503.14721},
      archivePrefix={arXiv},
      primaryClass={cond-mat.supr-con},
      url={https://arxiv.org/abs/2503.14721}, 
}

@ARTICLE{Wang2025,
  title     = "Density matrix renormalization group study of the
               quantum-geometry-facilitated pair density wave order",
  author    = "Wang, Hao-Xin and Huang, Wen",
  journal   = "Sci. China Phys. Mech. Astron.",
  publisher = "Springer Science and Business Media LLC",
  volume    =  68,
  number    =  9,
  pages     = "1--7",
  month     =  sep,
  year      =  2025,
  URL = "https://link.springer.com/article/10.1007/s11433-025-2701-1"
}

@misc{Prijon2025,
      title={Superfluid stiffness of superconductors with delicate topology}, 
      author={Tijan Prijon and Sebastian D. Huber and Kukka-Emilia Huhtinen},
      year={2025},
      eprint={2507.16909},
      archivePrefix={arXiv},
      primaryClass={cond-mat.supr-con},
      url={https://arxiv.org/abs/2507.16909}, 
}

@ARTICLE{Kitamura2021,
  title     = "Thermodynamic electric quadrupole moments of nematic phases from
               first-principles calculations",
  author    = "Kitamura, Taisei and Ishizuka, Jun and Daido, Akito and Yanase,
               Youichi",
  journal   = "Phys. Rev. B Condens. Matter",
  publisher = "American Physical Society",
  volume    =  103,
  number    =  24,
  pages     =  245114,
  month     =  jun,
  year      =  2021,
  URL = "https://journals.aps.org/prb/abstract/10.1103/PhysRevB.103.245114"
}

@ARTICLE{Berezinskii1971,
  title     = "Destruction of long-range order in one-dimensional and
               two-dimensional systems having a continuous symmetry group {I}.
               Classical systems",
  author    = "Berezinskii, V L",
  journal   = "Sov. Phys. JETP",
  publisher = "jetp.ac.ru",
  volume    =  32,
  number    =  3,
  pages     = "493--500",
  year      =  1971
}

@ARTICLE{Kosterlitz1973,
  title     = "Ordering, metastability and phase transitions in two-dimensional
               systems",
  author    = "Kosterlitz, J M and Thouless, D J",
  journal   = "J. Phys. C: Solid State Phys.",
  publisher = "IOP Publishing",
  volume    =  6,
  number    =  7,
  pages     =  1181,
  month     =  apr,
  year      =  1973,
  URL = "https://iopscience.iop.org/article/10.1088/0022-3719/6/7/010"
}

@BOOK{Mahan2010,
  title     = "Many-particle physics",
  author    = "Mahan, Gerald D",
  publisher = "Springer",
  address   = "Berlin",
  series    = "Physics of Solids and Liquids",
  month     =  dec,
  year      =  2010,
  URL = "https://link.springer.com/book/10.1007/978-1-4757-5714-9"
}

@INCOLLECTION{Liu2024,
  title     = "Recent developments in fractional Chern insulators",
  author    = "Liu, Zhao and Bergholtz, Emil J",
  booktitle = "Encyclopedia of Condensed Matter Physics",
  publisher = "Elsevier",
  pages     = "515--538",
  month     =  jan,
  year      =  2024,
  URL = "https://www.sciencedirect.com/science/article/pii/B9780323908009001360?via%3Dihub"
}

@ARTICLE{Park2021,
  title     = "Tunable strongly coupled superconductivity in magic-angle twisted
               trilayer graphene",
  author    = "Park, Jeong Min and Cao, Yuan and Watanabe, Kenji and Taniguchi,
               Takashi and Jarillo-Herrero, Pablo",
  journal   = "Nature",
  publisher = "Springer Science and Business Media LLC",
  volume    =  590,
  number    =  7845,
  pages     = "249--255",
  month     =  feb,
  year      =  2021,
  URL = "https://www.nature.com/articles/s41586-021-03192-0"
}

@ARTICLE{Hao2021,
  title     = "Electric field-tunable superconductivity in alternating-twist
               magic-angle trilayer graphene",
  author    = "Hao, Zeyu and Zimmerman, A M and Ledwith, Patrick and Khalaf,
               Eslam and Najafabadi, Danial Haie and Watanabe, Kenji and
               Taniguchi, Takashi and Vishwanath, Ashvin and Kim, Philip",
  journal   = "Science",
  publisher = "American Association for the Advancement of Science (AAAS)",
  volume    =  371,
  number    =  6534,
  pages     = "1133--1138",
  month     =  mar,
  year      =  2021,
  URL = "https://www.science.org/doi/10.1126/science.abg0399"
}

@ARTICLE{Hardy1993-sq,
  title     = "Precision measurements of the temperature dependence of lambda in
               {YBa$_2$Cu$_3$O$_{6.95}$}: Strong evidence for nodes in the gap function",
  author    = "Hardy, W N and Bonn, D A and Morgan, D C and Liang, R and Zhang,
               K",
  journal   = "Phys. Rev. Lett.",
  publisher = "American Physical Society (APS)",
  volume    =  70,
  number    =  25,
  pages     = "3999--4002",
  month     =  "21~" # jun,
  year      =  1993,
  url       = "http://link.aps.org/pdf/10.1103/PhysRevLett.70.3999",
  doi       = "10.1103/PhysRevLett.70.3999"
}

@ARTICLE{Huhtinen2022-mu,
  title     = "Revisiting flat band superconductivity: Dependence on minimal
               quantum metric and band touchings",
  author    = "Huhtinen, Kukka-Emilia and Herzog-Arbeitman, Jonah and Chew,
               Aaron and Bernevig, Bogdan A and T{\"{o}}rm{\"{a}}, P{\"{a}}ivi",
  journal   = "Phys. Rev. B",
  publisher = "American Physical Society",
  volume    =  106,
  number    =  1,
  pages     =  014518,
  month     =  "26~" # jul,
  year      =  2022,
  url       = "https://link.aps.org/doi/10.1103/PhysRevB.106.014518",
  doi       = "10.1103/PhysRevB.106.014518"
}

@ARTICLE{Marzari1997,
  title     = "Maximally localized generalized {W}annier functions for composite
               energy bands",
  author    = "Marzari, Nicola and Vanderbilt, David",
  journal   = "Phys. Rev. B Condens. Matter",
  publisher = "American Physical Society",
  volume    =  56,
  number    =  20,
  pages     = "12847--12865",
  month     =  nov,
  year      =  1997,
  URL = "https://journals.aps.org/prb/abstract/10.1103/PhysRevB.56.12847"
}

@ARTICLE{Vergniory2019,
  title    = "A complete catalogue of high-quality topological materials",
  author   = "Vergniory, M G and Elcoro, L and Felser, Claudia and Regnault,
              Nicolas and Bernevig, B Andrei and Wang, Zhijun",
  journal  = "Nature",
  volume   =  566,
  number   =  7745,
  pages    = "480--485",
  month    =  feb,
  year     =  2019,
  URL = "https://www.nature.com/articles/s41586-019-0954-4"
}

@ARTICLE{Zhang2019,
  title    = "Catalogue of topological electronic materials",
  author   = "Zhang, Tiantian and Jiang, Yi and Song, Zhida and Huang, He and
              He, Yuqing and Fang, Zhong and Weng, Hongming and Fang, Chen",
  journal  = "Nature",
  volume   =  566,
  number   =  7745,
  pages    = "475--479",
  month    =  feb,
  year     =  2019,
  URL = "https://www.nature.com/articles/s41586-019-0944-6"
}

@ARTICLE{Yu2023,
  title     = "Euler-obstructed nematic nodal superconductivity in twisted
               bilayer graphene",
  author    = "Yu, Jiabin and Xie, Ming and Wu, Fengcheng and Das Sarma, Sankar",
  journal   = "Phys. Rev. B Condens. Matter",
  publisher = "American Physical Society",
  volume    =  107,
  number    =  20,
  pages     = "L201106",
  month     =  may,
  year      =  2023,
  URL = "https://doi.org/10.1103/PhysRevB.107.L201106"
}

@ARTICLE{Provost1980,
  title    = "Riemannian structure on manifolds of quantum states",
  author   = "Provost, J P and Vallee, G",
  journal  = "Commun. Math. Phys.",
  volume   =  76,
  number   =  3,
  pages    = "289--301",
  month    =  sep,
  year     =  1980,
URL = "https://link.springer.com/article/10.1007/BF02193559"
}

@ARTICLE{Mazin2008,
  title     = "{Unconventional Superconductivity with a Sign Reversal in the
               Order Parameter of} {LaFeAsO}$_{1-x}${F}$_{x}$",
  author    = "Mazin, I I and Singh, D J and Johannes, M D and Du, M H",
  journal   = "Phys. Rev. Lett.",
  publisher = "American Physical Society",
  volume    =  101,
  number    =  5,
  pages     =  057003,
  month     =  jul,
  year      =  2008,
  URL = "https://journals.aps.org/prl/abstract/10.1103/PhysRevLett.101.057003"
}

@ARTICLE{Kuroki2008,
  title     = "Unconventional {P}airing {O}riginating from the {D}isconnected Fermi
               {S}urfaces of {S}uperconducting {{L}a{F}e{A}s{O}}$_{1\ensuremath{-}x}${F}$_{x}$",
  author    = "Kuroki, K and Onari, S and Arita, R and Usui, H and Tanaka, Y and
               {others}",
  journal   = "Phys. Rev. Lett.",
  publisher = "APS",
  year      =  2008,
  URL = "https://journals.aps.org/prl/abstract/10.1103/PhysRevLett.101.087004"
}

@ARTICLE{Sodemann2015,
  title    = "Quantum {N}onlinear {H}all {E}ffect {I}nduced by {B}erry {C}urvature {D}ipole in
              {T}ime-{R}eversal {I}nvariant {M}aterials",
  author   = "Sodemann, Inti and Fu, Liang",
  journal  = "Phys. Rev. Lett.",
  volume   =  115,
  number   =  21,
  pages    =  216806,
  month    =  nov,
  year     =  2015,
  URL = "https://journals.aps.org/prl/abstract/10.1103/PhysRevLett.115.216806"
}

@ARTICLE{Nagaosa2010,
  title     = "Anomalous {H}all effect",
  author    = "Nagaosa, Naoto and Sinova, Jairo and Onoda, Shigeki and
               MacDonald, A H and Ong, N P",
  journal   = "Rev. Mod. Phys.",
  publisher = "American Physical Society",
  volume    =  82,
  number    =  2,
  pages     = "1539--1592",
  month     =  may,
  year      =  2010,
  URL = "https://journals.aps.org/rmp/abstract/10.1103/RevModPhys.82.1539"
}

@ARTICLE{Nagaosa2017,
  title    = "Concept of {Q}uantum {G}eometry in {O}ptoelectronic Processes in {S}olids:
              {A}pplication to {S}olar {C}ells",
  author   = "Nagaosa, Naoto and Morimoto, Takahiro",
  journal  = "Adv. Mater.",
  volume   =  29,
  number   =  25,
  month    =  jul,
  year     =  2017,
  URL = "https://advanced.onlinelibrary.wiley.com/doi/10.1002/adma.201603345"
}

@ARTICLE{Daido2024,
  title     = "Quantum geometry encoded to pair potentials",
  author    = "Daido, Akito and Kitamura, Taisei and Yanase, Youichi",
  journal   = "Phys. Rev. B.",
  publisher = "American Physical Society (APS)",
  volume    =  110,
  number    =  9,
  pages     =  094505,
  month     =  sep,
  year      =  2024,
  URL = "https://journals.aps.org/prb/abstract/10.1103/PhysRevB.110.094505"
}

@misc{Oh2025,
      title={Magnetic phase transitions driven by quantum geometry}, 
      author={Oh, Chang-geun and Kitamura, Taisei and Daido,Akito and Rhim, Jun-Won and Yanase, Youichi},
      year={2025},
      eprint={2509.13618},
      archivePrefix={arXiv},
      primaryClass={cond-mat.str-el},
      url={https://arxiv.org/abs/2509.13618}, 
}

@ARTICLE{Cao2018,
  title    = "Unconventional superconductivity in magic-angle graphene
              superlattices",
  author   = "Cao, Yuan and Fatemi, Valla and Fang, Shiang and Watanabe, Kenji
              and Taniguchi, Takashi and Kaxiras, Efthimios and Jarillo-Herrero,
              Pablo",
  journal  = "Nature",
  volume   =  556,
  number   =  7699,
  pages    = "43--50",
  month    =  apr,
  year     =  2018,
  URL = "https://www.nature.com/articles/nature26160"
}

@ARTICLE{Mielke1992,
  title     = "{Exact ground states for the Hubbard model on the Kagome lattice}",
  author    = "Mielke, A",
  journal   = "J. Phys. A Math. Gen.",
  publisher = "IOP Publishing",
  volume    =  25,
  number    =  16,
  pages     = "4335--4345",
  month     =  aug,
  year      =  1992,
  URL = "https://www.semanticscholar.org/paper/Ferromagnetic-ground-states-for-the-Hubbard-model-Mielke/ae6c98c222bb0664816bbf226215bff46b2b3f01"
}

@ARTICLE{Mielke1991,
  title     = "{Ferromagnetic ground states for the Hubbard model on line graphs}",
  author    = "Mielke, A",
  journal   = "J. Phys. A Math. Gen.",
  publisher = "IOP Publishing",
  volume    =  24,
  number    =  2,
  pages     = "L73--L77",
  month     =  jan,
  year      =  1991,
  URL = "https://www.semanticscholar.org/paper/Ferromagnetic-ground-states-for-the-Hubbard-model-Mielke/ae6c98c222bb0664816bbf226215bff46b2b3f01"
}

@ARTICLE{Herzog2022prl,
  title     = "{Superfluid Weight Bounds from Symmetry and Quantum Geometry in Flat Bands}",
  author    = "Herzog-Arbeitman, Jonah and Peri, Valerio and Schindler, Frank
               and Huber, Sebastian D and Bernevig, B Andrei",
  journal   = "Phys. Rev. Lett.",
  publisher = "American Physical Society (APS)",
  volume    =  128,
  number    =  8,
  pages     =  087002,
  month     =  feb,
  year      =  2022,
  URL = "https://journals.aps.org/prl/abstract/10.1103/PhysRevLett.128.087002"
}

@misc{Kudo2025,
      title={Odd-parity magnetism by quantum geometry}, 
      author={Kanta Kudo and Youichi Yanase},
      year={2025},
      eprint={2505.20907},
      archivePrefix={arXiv},
      primaryClass={cond-mat.str-el},
      url={https://arxiv.org/abs/2505.20907}, 
}

@ARTICLE{Kitamura2022superconductivity,
  title     = "Superconductivity in monolayer {FeSe} enhanced by quantum
               geometry",
  author    = "Kitamura, Taisei and Yamashita, Tatsuya and Ishizuka, Jun and
               Daido, Akito and Yanase, Youichi",
  journal   = "Phys. Rev. Research",
  publisher = "American Physical Society",
  volume    =  4,
  number    =  2,
  pages     =  023232,
  month     =  jun,
  year      =  2022,
  URL = "https://journals.aps.org/prresearch/abstract/10.1103/PhysRevResearch.4.023232"
}

@misc{Kitamura2025,
      title={Quantum geometric ferromagnetism by singular saddle point}, 
      author={Taisei Kitamura and Hiroki Nakai and Akito Daido and Youichi Yanase},
      year={2025},
      eprint={2505.01089},
      archivePrefix={arXiv},
      primaryClass={cond-mat.str-el},
      url={https://arxiv.org/abs/2505.01089}, 
}

@misc{Hirobe2025,
      title={{Anomalous Temperature Dependence of Quantum-Geometric Superfluid Weight}}, 
      author={Yuma Hirobe and Taisei Kitamura and Youichi Yanase},
      year={2025},
      eprint={2505.13065},
      archivePrefix={arXiv},
      primaryClass={cond-mat.supr-con},
      url={https://arxiv.org/abs/2505.13065}, 
}

@ARTICLE{Xie2020,
  title     = "Topology-{B}ounded {S}uperfluid {W}eight in {T}wisted {B}ilayer {G}raphene",
  author    = "Xie, Fang and Song, Zhida and Lian, Biao and Bernevig, B Andrei",
  journal   = "Phys. Rev. Lett.",
  publisher = "APS",
  volume    =  124,
  number    =  16,
  pages     =  167002,
  month     =  apr,
  year      =  2020,
  URL = "https://doi.org/10.1103/PhysRevLett.124.167002"
}

@ARTICLE{Hu2019,
  title    = "Geometric and {C}onventional {C}ontribution to the {S}uperfluid {W}eight
              in {T}wisted {B}ilayer {G}raphene",
  author   = "Hu, Xiang and Hyart, Timo and Pikulin, Dmitry I and Rossi, Enrico",
  journal  = "Phys. Rev. Lett.",
  volume   =  123,
  number   =  23,
  pages    =  237002,
  month    =  dec,
  year     =  2019,
  URL = "https://journals.aps.org/prl/abstract/10.1103/PhysRevLett.123.237002"
}

@ARTICLE{Julku2020,
  title     = "Superfluid weight and {Berezinskii}-{Kosterlitz}-{Thouless} transition
               temperature of twisted bilayer graphene",
  author    = "Julku, A and Peltonen, T J and Liang, L and Heikkilä, T T and
               Törmä, P",
  journal   = "Phys. Rev. B: Condens. Matter Mater. Phys.",
  publisher = "APS",
  year      =  2020,
  URL = "https://journals.aps.org/prb/abstract/10.1103/PhysRevB.101.060505"
}

@ARTICLE{Tanaka2025,
  title     = "Superfluid stiffness of magic-angle twisted bilayer graphene",
  author    = "Tanaka, Miuko and Wang, Joel \^{I}-j and Dinh, Thao H and
               Rodan-Legrain, Daniel and Zaman, Sameia and Hays, Max and
               Almanakly, Aziza and Kannan, Bharath and Kim, David K and
               Niedzielski, Bethany M and Serniak, Kyle and Schwartz, Mollie E
               and Watanabe, Kenji and Taniguchi, Takashi and Orlando, Terry P
               and Gustavsson, Simon and Grover, Jeffrey A and Jarillo-Herrero,
               Pablo and Oliver, William D",
  journal   = "Nature",
  publisher = "Springer Science and Business Media LLC",
  volume    =  638,
  number    =  8049,
  pages     = "99--105",
  month     =  feb,
  year      =  2025,
  URL = "https://www.nature.com/articles/s41586-024-08494-7"
}

@ARTICLE{Banerjee2025,
  title     = "Superfluid stiffness of twisted trilayer graphene superconductors",
  author    = "Banerjee, Abhishek and Hao, Zeyu and Kreidel, Mary and Ledwith,
               Patrick and Phinney, Isabelle and Park, Jeong Min and Zimmerman,
               Andrew and Wesson, Marie E and Watanabe, Kenji and Taniguchi,
               Takashi and Westervelt, Robert M and Yacoby, Amir and
               Jarillo-Herrero, Pablo and Volkov, Pavel A and Vishwanath, Ashvin
               and Fong, Kin Chung and Kim, Philip",
  journal   = "Nature",
  publisher = "Springer Science and Business Media LLC",
  volume    =  638,
  number    =  8049,
  pages     = "93--98",
  month     =  feb,
  year      =  2025,
  URL = "https://www.nature.com/articles/s41586-024-08444-3"
}

@ARTICLE{Tian2023,
  title    = "{Evidence for Dirac flat band superconductivity enabled by quantum
              geometry}",
  author   = "Tian, Haidong and Gao, Xueshi and Zhang, Yuxin and Che, Shi and
              Xu, Tianyi and Cheung, Patrick and Watanabe, Kenji and Taniguchi,
              Takashi and Randeria, Mohit and Zhang, Fan and Lau, Chun Ning and
              Bockrath, Marc W",
  journal  = "Nature",
  volume   =  614,
  number   =  7948,
  pages    = "440--444",
  month    =  feb,
  year     =  2023,
  URL = "https://www.nature.com/articles/s41586-022-05576-2"
}

@misc{Peotta2023,
      title={Quantum geometry in superfluidity and superconductivity}, 
      author={Sebastiano Peotta and Kukka-Emilia Huhtinen and Päivi Törmä},
      year={2023},
      eprint={2308.08248},
      archivePrefix={arXiv},
      primaryClass={cond-mat.quant-gas},
      url={https://arxiv.org/abs/2308.08248}, 
}

@ARTICLE{Liang2017,
  title     = "Band geometry, {B}erry curvature, and superfluid weight",
  author    = "Liang, Long and Vanhala, Tuomas I and Peotta, Sebastiano and
               Siro, Topi and Harju, Ari and Törmä, Päivi",
  journal   = "Phys. Rev. B Condens. Matter",
  publisher = "American Physical Society",
  volume    =  95,
  number    =  2,
  pages     =  024515,
  month     =  jan,
  year      =  2017,
  URL = "https://doi.org/10.1103/PhysRevB.95.024515"
}

@ARTICLE{Peotta2015,
  title    = "Superfluidity in topologically nontrivial flat bands",
  author   = "Peotta, Sebastiano and Törmä, Päivi",
  journal  = "Nat. Commun.",
  volume   =  6,
  number   =  1,
  pages    =  8944,
  month    =  nov,
  year     =  2015,
  URL = "https://www.nature.com/articles/ncomms9944"
}

@ARTICLE{Nuckolls2024,
  title     = "A microscopic perspective on moiré materials",
  author    = "Nuckolls, Kevin P and Yazdani, Ali",
  journal   = "Nat. Rev. Mater.",
  publisher = "Springer Science and Business Media LLC",
  volume    =  9,
  number    =  7,
  pages     = "460--480",
  month     =  may,
  year      =  2024,
  URL = "https://www.nature.com/articles/s41578-024-00682-1"
}

@ARTICLE{Andrei2021,
  title     = "The marvels of moiré materials",
  author    = "Andrei, Eva Y and Efetov, Dmitri K and Jarillo-Herrero, Pablo and
               MacDonald, Allan H and Mak, Kin Fai and Senthil, T and Tutuc,
               Emanuel and Yazdani, Ali and Young, Andrea F",
  journal   = "Nature Reviews Materials",
  publisher = "Nature Publishing Group",
  volume    =  6,
  number    =  3,
  pages     = "201--206",
  month     =  mar,
  year      =  2021,
  URL = "https://www.nature.com/articles/s41578-021-00284-1"
}

@misc{Zhang2025,
      title={Identifying Instabilities with Quantum Geometry in Flat Band Systems}, 
      author={Jia-Xin Zhang and Wen O. Wang and Leon Balents and Lucile Savary},
      year={2025},
      eprint={2504.03882},
      archivePrefix={arXiv},
      primaryClass={cond-mat.str-el},
      url={https://arxiv.org/abs/2504.03882}, 
}

@ARTICLE{Jujo2001,
  title     = "Fermi Liquid Theory on Transport Phenomena in the Superconducting
               State",
  author    = "Jujo, Takanobu",
  journal   = "J. Phys. Soc. Jpn.",
  publisher = "The Physical Society of Japan",
  volume    =  70,
  number    =  5,
  pages     = "1349--1363",
  month     =  may,
  year      =  2001,
  URL = "https://journals.jps.jp/doi/10.1143/JPSJ.70.1349"
}

@ARTICLE{Scalapino1992,
  title     = "Superfluid density and the {D}rude weight of the {H}ubbard model",
  author    = "Scalapino, D J and White, S R and Zhang, S C",
  journal   = "Phys. Rev. Lett.",
  publisher = "American Physical Society (APS)",
  volume    =  68,
  number    =  18,
  pages     = "2830--2833",
  month     =  may,
  year      =  1992,
  URL = "https://journals.aps.org/prl/abstract/10.1103/PhysRevLett.68.2830"
}

@ARTICLE{Scalapino1993,
  title     = "Insulator, metal, or superconductor: The criteria",
  author    = "Scalapino, D J and White, S R and Zhang, S",
  journal   = "Phys. Rev. B Condens. Matter",
  publisher = "American Physical Society (APS)",
  volume    =  47,
  number    =  13,
  pages     = "7995--8007",
  month     =  apr,
  year      =  1993,
  URL = "https://journals.aps.org/prb/abstract/10.1103/PhysRevB.47.7995"
}

@book{Altland2010, place={Cambridge}, edition={2}, title={Condensed Matter Field Theory}, publisher={Cambridge University Press}, author={Altland, Alexander and Simons, Ben D.}, year={2010},url = "https://www.cambridge.org/core/books/condensed-matter-field-theory/0A8DE6503ED868D96985D9E7847C63FF"}

@ARTICLE{Jiang2023,
  title    = "{Pair Density Waves from Local Band Geometry}",
  author   = "Jiang, Guodong and Barlas, Yafis",
  journal  = "Phys. Rev. Lett.",
  volume   =  131,
  number   =  1,
  pages    =  016002,
  month    =  jul,
  year     =  2023,
  URL = "https://journals.aps.org/prl/abstract/10.1103/PhysRevLett.131.016002"
}

@ARTICLE{Kitamura2022,
  title     = "{Quantum geometric effect on Fulde-Ferrell-Larkin-Ovchinnikov
               superconductivity}",
  author    = "Kitamura, Taisei and Daido, Akito and Yanase, Youichi",
  journal   = "Phys. Rev. B Condens. Matter",
  publisher = "American Physical Society",
  volume    =  106,
  number    =  18,
  pages     =  184507,
  month     =  nov,
  year      =  2022,
  URL = "https://journals.aps.org/prb/abstract/10.1103/PhysRevB.106.184507"
}

@ARTICLE{Chen2023,
  title    = "{Pair density wave facilitated by Bloch quantum geometry in nearly
              flat band multiorbital superconductors}",
  author   = "Chen, Weipeng and Huang, Wen",
  journal  = "Sci. China. Ser. G: Phys. Mech. Astron.",
  volume   =  66,
  number   =  8,
  pages    =  287212,
  month    =  jul,
  year     =  2023,
 URL = "https://link.springer.com/article/10.1007/s11433-023-2122-4"
}

@ARTICLE{Kitamura2023,
  title     = "Quantum geometry induced anapole superconductivity",
  author    = "Kitamura, Taisei and Kanasugi, Shota and Chazono, Michiya and
               Yanase, Youichi",
  journal   = "Phys. Rev. B Condens. Matter",
  publisher = "American Physical Society",
  volume    =  107,
  number    =  21,
  pages     =  214513,
  month     =  jun,
  year      =  2023,
 URL = "https://journals.aps.org/prb/abstract/10.1103/PhysRevB.107.214513"
}

\end{document}